# Mapping Phonon Polaritons with Visible Light

Kiernan E. Arledge[1,2,*], Chase T. Ellis[3], Nazli Rasouli Sarabi[2,4], Vincent R. Whiteside[1,2], Chul Soo Kim[3], Mijin Kim[3], Daniel C. Ratchford[3], Michael A. Meeker[3,5], Binbin Weng[2,4], and Joseph G. Tischler[1,2,*]

[1]Homer L. Dodge Department of Physics and Astronomy, University of Oklahoma, Norman, 73019, USA
[2] Center for Quantum Research and Technologies, University of Oklahoma, Norman, 73019, USA
[3]U.S. Naval Research Laboratory, Washington D.C., 20375, USA
[4]School of Electrical and Computer Engineering, University of Oklahoma, Norman, 73019, USA
[5]Advanced Research Center, City University of New York, New York, 10017, USA
*

[*]email: tischler@ou.edu and kiernan.e.arledge-1@ou

## ABSTRACT

Phonon polaritons (PhPs) are hybrid photon-phonon waves which enable strong light-matter interactions and subdiffractional confinement, potentially empowering applications in sensing, nonlinear optics and nanoscale energy manipulation. In this work, we use confocal Raman microscopy to investigate the coupling between bulk phonon modes and localized surface phonon polariton (SPhP) modes in indium phosphide (InP) nanopillars and 4H-silicon carbide (4H-SiC) gratings. The Raman intensity within the nanostructures is described in terms of the SPhP eigenmodes and used to reconstruct the field intensity, providing a method to map SPhP eigenmodes using visible and near-IR light. Our results indicate that, contrary to expectation, all Raman-active bulk phonon modes of InP and 4H-SiC couple to the localized SPhP modes. Further, we confirm that polarizability selection rules form the predominant coupling mechanism between phonons and SPhP modes, with electron-phonon coupling playing a role for certain phonon modes ($A_1$(LO) and $E_1$(TO) in 4H-SiC). These observations provide a method for extending Raman studies of PhP modes to achieve full 3D reconstruction of the PhP eigenmodes and visualize light-matter interactions within nanostructures, thus advancing Raman scattering as a technique for understanding PhP modes.

## Introduction

Phonon polaritons (PhPs) are electromagnetic modes that emerge when photons couple with optical phonons in a polar dielectric material. PhPs often come in the form of electromagnetic fields concentrated at the crystal surface, where they are termed surface phonon polaritons (SPhPs). SPhPs have been shown to confine electromagnetic fields far below the diffraction limit analogous to surface plasmon polaritons (SPPs), but without the strong optical losses associated with SPP modes[1,2]. Furthermore, PhP modes are also spectrally tunable through tailoring nanostructure geometry[3], introducing free carriers[4,5], and exploiting coupling to plasmons[6], zone-folded phonons[7], or other solid state excitations[8–10]. These advantages have been leveraged to demonstrate the potential for PhPs in surface-enhanced sensing[11], nanophotonic circuits[12], super-resolution imaging[13], and nonlinear optics[14,15].

The tunable and low-loss nature of PhP resonators also makes them an appealing platform for developing chip-scale mid-infrared (mid-IR) emitters[16,17]. Efforts to demonstrate the viability of an electrically-pumped PhP-based emitter have been impeded by the presumed weak interaction between phonons and PhP modes in bulk materials[17,18]. Recent investigations have suggested that this limitation can be overcome via coupling of PhP modes to zone-folded longitudinal optical (ZFLO) phonons in 4H-Silicon Carbide (SiC)[4], although engineering such coupling places constraints on material selection and nanostructure morphology[7].

Assessing the promise of PhP modes as mid-IR emitters clearly requires an investigation of the interactions between electrons, phonons, and PhP modes. Studies on PhP-ZFLO[7] and PhP-electron[4,19] interactions have been very important in this regard however, full explorations of bulk phonon-PhP-electron coupling have remained limited. Practical challenges have played a role in this delay. Investigations of PhPs in dielectric nanostructures have increasingly revealed strong spatial dependence on light-matter interactions - for example, the strength of electron-phonon coupling in many nanostructures depends on the induced 3D charge distribution[4,19,20]. Therefore, any investigation into light-matter coupling would ideally demonstrate how the full 3D PhP eigenmodes interact with solid state excitations to be complete.

To date, investigating PhP eigenmodes has primarily been accomplished using near-field IR microscopy, such as near-field scanning optical microscopy (NSOM)[21] and Nano-FTIR techniques[22]. Recently, monochromatic electron energy-loss spectroscopy (EELS) in scanning transmission electron microscope (STEM) setups has also been employed to visualize SPhP modes in nanostructured materials[23,24]. This work was extended to a tomographic setup combining tilted EELS spectra with reconstruction algorithms in magnesium oxide nanostructures to first achieve a 3D mapping of SPhP eigenmodes[25].

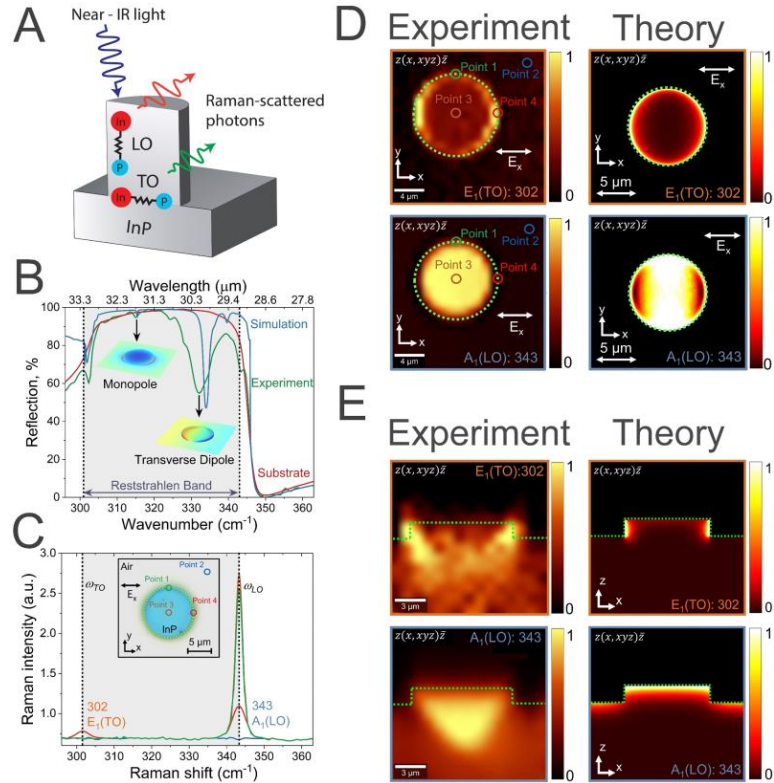

**Figure 1. Phonon-SPhP coupling in InP nanopillars.** (**A**) Illustration of confocal Raman measurements on InP pillars. Near-IR light focused by the microscope objective produces Raman scattering from longitudinal and transverse optical phonons (LO and TO). (**B**) FTIR reflection spectrum of a periodic array of InP pillars (green curve) compared to the simulated spectrum obtained using COMSOL (blue curve) and the FTIR reflection of the InP substrate (red curve). Insets: Charge distribution of monopole and transverse dipole resonances. (**C**) Unpolarized, backscattering micro-Raman spectra measured at 4 different points (as shown in the figure inset, taken at a focus position roughly 0.5 μm above the pillar-substrate interface) across the top face an InP pillar, showing the spatial dependence of the LO and TO phonon modes, which bound the Reststrahlen band (gray shaded region) in (**B**). (**D**) (left) Experimentally measured and (right) theoretically simulated Raman intensity maps obtained for the 302 cm$^{-1}$ $E_1$(TO) (upper panels) and 343 cm$^{-1}$ $A_1$(LO) (lower panels) phonon modes for an x-y cross-sectional view, showing the distinct localization of Raman intensity for each phonon mode. The dotted green lines represent the InP-air interface. (**E**) Same as (**D**) for a y-z cross-sectional view.

Increasingly, these efforts have been augmented with the techniques of Raman spectroscopy, leveraging its ability to provide abundant information on light-matter interactions. Raman techniques have particularly been used to probe PhPs in 2D materials. Initial works demonstrated resonant Raman spectroscopy as a valuable tool to probe PhP modes at Van der Waals (VdW) material interfaces, revealing effects such as Raman activity of silent phonons[26], enhancement of PhP Raman features[27], and PhPs outside the Reststrahlen band[28]. Further, recent investigations have shown that conventional backscattering Raman can probe PhP modes in 2D gallium selenide, utilizing the polarizability selection rules[29]. Such techniques are not restricted to 2D materials, micro-Raman investigations of 4H-SiC nanopillars found that spatial dispersion of the permittivity in dielectric nanostructures can play a major role in the spectral behavior of PhPs[20]. This discovery highlights the fact that much of the physics related to the spatially dependent interactions of electrons, phonons, and SPhP modes in nanostructures remains uninvestigated. Therefore, efforts to understand the physics governing PhPs in nanostructures demand not only spectral probing of SPhP dispersion but also spatial mapping of light-matter interactions through the nanostructure volume with more sophisticated Raman techniques.

In this work, we report the mapping of SPhP electromagnetic fields in nanostructures of InP and 4H-SiC. We combined confocal Raman microscopy with an eigenmode reconstruction approach to provide a 3D view of localized polariton fields. Using this technique, we demonstrate that all Raman-active bulk phonon modes in both InP and 4H-SiC do couple to the localized surface polariton modes, in contrast to the assumption that the longitudinal polarization field of the LO phonon is non-interacting with the transverse (and thus orthogonal) SPhP modes. Our Raman maps confirm that the polarizability selection rules form the dominant mechanism through which SPhPs couple to phonon modes. For the specific case of the $A_1$(LO) and $E_1$(TO) phonon modes in 4H-SiC, electron-phonon coupling also plays a role, even for low-doped substrates. These results challenge the conventional understanding of SPhP-phonon coupling and indicate that electrically pumped SPhP mid-IR emitters might be achievable with dielectric nanostructures. Furthermore, our work provides a method for real-space mapping of SPhP eigenmodes in a manner analogous to tomographic EELS, but with no destructive sample preparation and not requiring the specialized equipment of highly monochromatic EELS. Thus, Raman imaging can be a complimentary tool for the visualizing of SPhP modes which also provides a natural platform for investigating the interactions between electrons, phonons, SPhPs, and other solid-state excitations.

## Results

### Coupling between phonons and SPhPs – InP pillars

To investigate the correlation between Raman measurements and the near-field structure of the SPhP modes, we perform IR reflectance and scanning confocal Raman mapping measurements on arrays of cylindrical InP pillars with a radius of 10 µm and height of 1.5 µm, as depicted in Fig. 1A (see methods measurement and sample details). To avoid photodoping and inadvertently phototuning the SPhP resonances while performing Raman measurements[4], we probe the Raman spectra by employing a below-gap, near-IR laser centered at λ = 1064 nm.

As shown in Figure 1B, the measured IR reflectance spectrum for the InP pillar array (green curve) deviates significantly from measurements of the bare substrate (red curve), indicating the excitation of localized SPhP modes. These measurements are in excellent agreement with full-wave electromagnetic simulations (blue line). The InP pillars support two primary localized SPhP resonant modes superimposed on the highly reflective Reststrahlen band indicated by gray shading. Numerical calculations of the surface charge distribution at resonance indicate that the higher-energy resonance is a transverse dipole mode, consisting of charges localized at opposite ends of the pillar (see inset Figure 1B). The lower-energy resonance is a monopole mode, consisting of charge distributed across the pillar face (see inset Figure 1B). These mode profiles are similar to SPhP resonances observed in other nanopillar systems[1,4]. The spatial profiles of these modes localized in the 10 µm radii pillars with an excitation wavelength of ~30 µm indicates the subdiffractional nature of the SPhP modes and resonances at pillar radii below 1 µm are also observable (Fig. S3 in the Supporting Information).

By scanning the sample and recording a Raman spectrum at each scan position, we obtain Raman images that reveal the full 3D spatial map of the phonon modes within the nanostructure. Figure 1C shows a first investigation of the Raman modes in the pillars: several micro-Raman spectra collected in the x-y plane at a focus point roughly 0.5 µm above the pillar-substrate interface. All of the Raman scans in Fig. 1 were collected in the $z(x,xyz)\bar{z}$ geometry, where $z$ and $\bar{z}$ on either side of the parentheses denote the propagation direction of the incident and scattered light, $x$ denotes the polarization of the incident laser, and $xyz$ denotes the polarization of light collected for the detector (unpolarized, in this case). A representative $z(x,xyz)\bar{z}$ Raman spectrum for InP (Fig. 1C, brown line corresponding to Point 3 at the pillar center) reveals only two phonon modes: $E_1$(TO) and $A_1$(LO), which bound the Reststrahlen band (Fig. 1B), as expected for a polar zinc blende crystal such as bulk InP[30]. As shown in Fig. 1C, distinct spectral responses are extracted at different positions on the pillar face. In particular, at point 1 on the top edge of the pillar (green curve), the TO phonon has much stronger Raman scattering, and the LO phonon scattering becomes suppressed. This dramatic change in the spectral intensity indicates that the Raman scattering by phonons demonstrates remarkable spatial dependence in the nanostructure geometry, in contrast to the uniform Raman intensity of a bulk material.

To further understand the phonon scattering behavior, we show Raman spectral images in the same "top-down view" (x-y plane) at the same focus position on the z-axis as Fig. 1C, with the incident light polarized in the x-direction. Figure 1D shows the experimental (left panels) and simulated (right panels) intensity maps for the TO (upper panels) and LO (lower panels) phonon modes. The maps demonstrate that the $E_1$(TO) intensity is localized at the pillar edges and aligned with the laser polarization, similar to the SPhP transverse dipole mode (Fig. 1B inset). In contrast, the $A_1$(LO) mode Raman map reveals a nearly uniform scattering intensity for the LO mode across the pillar face, like the SPhP monopole mode (Fig. 1B inset). Images taken in the "cross-sectional views" (x-z plane) further support this conclusion, indicating that the scattering from the phonon modes are localized in a manner analogous to the SPhP modes. The Raman images provide evidence for our surprising finding that the TO and LO phonons interact with the distinct SPhP modes of the pillar, where the Raman scattering from the TO and LO phonon modes predominantly assume the field morphology of the dipole and monopole SPhP resonances, respectively.

**Field Mapping Theory**

To understand the SPhP response in the Raman maps, we demonstrate a basic, first-order reconstruction method to model the spatial distribution of the Raman intensity within the nanostructures. Our approach relies on the quasistatic approach, justified by the relatively large free-space wavelengths of the SPhPs on resonance, which are much larger than the pillar dimensions[31]. Following this assumption, we use a simple eigenmode reconstruction method to calculate the SPhP Raman response, inspired by EELS reconstructions of SPP and SPhP eigenmodes.

The basis for our approach comes from the polarizability selection rules, which have been used to calculate the Raman scattering intensities of propagating PhP modes in VdW materials[29]. We modify this approach to make it more suitable for describing localized SPhPs in nanostructures by expressing the Raman response of the nanostructure in terms of its resonant eigenmodes. Based on our observations in Fig. 1(B-D), we propose expressing the electric field which produces Raman scattering inside the nanostructure in terms of the SPhP eigenmodes, which have spatial distribution $\bar{u}_k(\bar{r})$; where $\bar{r}$ is the volume coordinate and $k$ indexes the eigenmode (e.g the monopole or dipole mode)[32,33]. This approach assumes that the electromagnetic fields of the mid-IR SPhP modes dominate the visible/near-IR Raman response via polarizability coupling. We will demonstrate that this approach provides a method to reconstruct the Raman intensity in the InP pillars, but is not expected to generally hold in situations when others effects (Mie resonances[34], nonlinear effects[35], other solid state excitations[36,37], etc.) have to be considered. Using these eigenmodes, we construct the Raman polarization ($\bar{P}_R$) induced in the structure as a result of Raman scattering[34,38]:

$$\bar{P}_R = \bar{\bar{\alpha}}^\sigma \cdot \bar{E}(\bar{r}) = C_k^\sigma \bar{\bar{\alpha}}^\sigma \bar{u}_k(\bar{r}) = \bar{P}_{R,k}^\sigma \tag{1}$$

where $\bar{\bar{\alpha}}^\sigma$ is the Raman polarizability tensor for phonon mode $\sigma$ (e.g $A_1$(LO) or $E_1$(TO) for InP), $\bar{E}(\bar{r})$ is the electric field inside the nanostructure (due to the excitation laser), and $C_k^\sigma$ is a coefficient which describes the coupling between the polarizability tensor of phonon mode $\bar{\bar{\alpha}}^\sigma$ and eigenmode $\bar{u}_k(\bar{r})$. The last equality emphasizes that Eq. (1) describes the Raman polarizability that is generated when an SPhP eigenmode couples with a phonon mode, imparting a spatial profile to the induced Raman polarizability.

While Eq. (1) contains the essential physics of the reconstruction process, it is the time-averaged Raman intensity that is measured. Using this approach, we write the time-averaged Raman intensity signal (in a volume element $dV$) in terms of its proportionality to the time-average of the Raman polarization amplitude

$$\langle I_{R,k}^\sigma \rangle \propto \langle |\bar{P}_{R,k}^\sigma| \rangle^2. \tag{2}$$

Equation 2 describes the Raman intensity for phonon mode $\sigma$ when coupled to SPhP mode $k$. Due to the assumed incoherent nature of the coupling between bulk phonons and SPhPs[34], we obtain the final expression for the Raman intensity of a given phonon mode by summing over the eigenmode basis:

$$\langle S_R^\sigma \rangle = \sum_k \langle I_{R,k}^\sigma \rangle \tag{3}$$

Equation 2 is similar to the expression for the Raman intensity of PhPs in 2D materials, where instead we have explicitly expressed the electric field in terms of the SPhP eigenmodes, which is the natural approach for localized polaritons in nanostructures. This representation also emphasizes the formal analogy to EELS reconstruction of SPhP modes, where Raman intensity maps can be used to visualize the SPhP eigenmodes comparable to EELS tomography of the local electromagnetic response driving light-matter interactions, the electromagnetic local density of states[25]. Our theoretical formalism is thus based on the notion that the induced Raman polarization (and hence Raman intensity) exhibits spatial dependence consistent with coupling to the polarization of the SPhP eigenmodes. Therefore, the measured Raman intensity for a given phonon mode in an InP pillar can be reconstructed as the combination of the SPhP eigenmodes subjected to the Raman selection rules for that phonon symmetry contained in $\bar{\bar{\alpha}}^\sigma$.

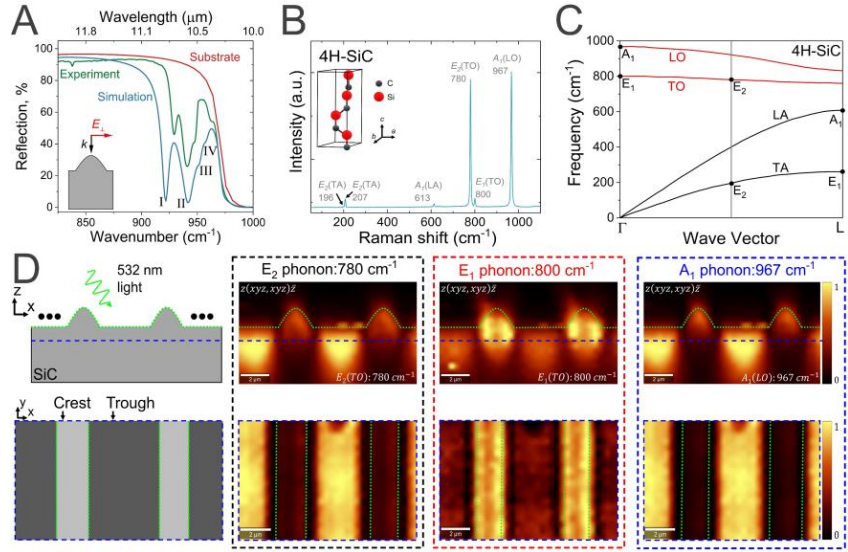

**Figure 2. 1D SiC gratings.** (**A**) FTIR reflection spectrum of SiC gratings (red curve) compared to the simulated spectrum obtained using COMSOL (green curve) and the reflection from the SiC substrate (black curve). Inset: illustration of the SiC grating unit cell. (**B**) Representative $z(x,x)\bar{z}$ Raman spectrum of 4H-SiC. All of the Raman-active phonon modes are labeled with their symmetry classification. Inset: Illustration of the 4H-SiC crystal structure. (**C**) Phonon dispersion curve of 4H-SiC based on Raman data. The dots denote all Raman-active phonon modes in the backscattering configuration. (**D**) (Top) x-z and (bottom) x-y unpolarized Raman maps of the three optical phonons in 4H-SiC: $E_2$(TO), $E_1$(TO), and $A_1$(LO). The dotted green lines show the SiC-air interface and the dotted blue lines denote the z-position of the x-y map (~ 1 μm below the SiC substrate).

We demonstrate that the spatially dependent Raman scattering originates in coupling to SPhPs by calculating the Raman polarization in the pillars. Full-wave electromagnetic simulations (COMSOL) are used to obtain the field distributions associated with the SPhP monopole and dipole modes. Then, using the Raman selection rules for each phonon mode, the Raman polarization for a given phonon mode is obtained via Eq. (1). The results of time-averaging the Raman polarization intensities and summing over the eigenmodes are presented in the right panels of Fig. 1D,E (further details on the reconstruction process are given in the supplementary information). The simulations reveal similar features to those seen in the experimental Raman maps, shown in the left panels. The simulated fields confirm that the Raman polarization for the TO and LO phonon modes predominantly display the morphology of the SPhP dipole and monopole resonances, respectively. This surprising discovery is not unreasonable given that the SPhP dipole is a transverse, in-plane mode and the TO phonon oscillates in-plane (and vice versa for the monopole mode and LO phonon). The Raman mapping of InP nanopillars indicates that the bulk phonons couple to the surface phonon polaritons through the material polarizability, enabling the 3D depth-profiling of SPhP modes in real space with a near-IR probe.

## Coupling Between Phonons and SPhPs - SiC

To further demonstrate the power of Raman mapping as a technique to investigate SPhP modes, we fabricated a 1D grating on 4H-SiC with period of 6 μm, linewidth of 3.3 μm, and height of 1.3 μm (see methods for details). The grating structure supports a variety of localized SPhP modes, as shown in the measured (red curve) and simulated (green curve) IR reflection spectrum of the nanostructure shown in Fig. 2A (a schematic of the unit cell is shown in the inset). In contrast to the symmetric InP pillar structure, the localized SPhP modes on the SiC grating cannot be simply described as monopole or dipole modes, and thus possess more complicated fields (Supplementary Fig. 6 shows the electric field distributions for all 4 modes). Nevertheless, we can take the set of SPhP resonances (which we denote as I,II,III, and IV in Fig. 2A) and use them as the basis for a more complicated Raman reconstruction.

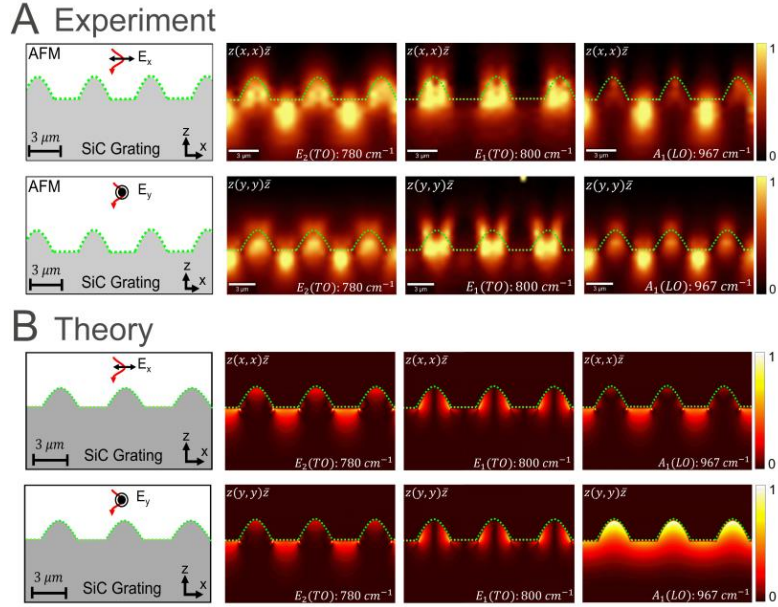

**Figure 3. Raman mapping of phonon-SPhP coupling: SiC gratings.** (**A**) (Left) Experimental non-contact AFM topography scans of the grating surface and (right) Raman intensity maps of the three optical phonon modes ($E_2$(TO), $E_1$(TO), $A_1$(LO)) for two geometric configurations ($z(x,x)\bar{z}$ and $z(y,y)\bar{z}$). The outline of the grating is highlighted with dotted green lines. (**B**) Same for full wave electromagnetic simulations of a 1D SiC grating.

Figure 2B shows a representative $z(x,x)\bar{z}$ Raman spectrum for 4H-SiC, whose crystal structure is shown in the inset. The wurtzite (hexagonal) structure of 4H-SiC supports 6 Raman-active phonon modes in the backscattering configuration: three acoustic phonons lie in the range 207 cm$^{-1}$ – 613 cm$^{-1}$, and three optical phonons lie in the Reststrahlen band (780 cm$^{-1}$ – 967 cm$^{-1}$), all of which fall into three symmetry classes: $E_2$, $E_1$, and $A_1$, as shown in the phonon dispersion of Fig. 2C.

As a first investigation of the Raman modes in the SiC gratings, we collect unpolarized $z(xyz,xyz)\bar{z}$ Raman maps, as presented in Fig. 2D. We consider two separate mapping configurations: a cross-sectional view in the x-z plane and a top-down view in the x-y plane (collected at a point ~ 1 μm below the SiC substrate), depicted with schematic illustrations in the left panel of Fig. 2D. The right three panels of Fig. 2D show the Raman maps for the three optical phonon modes: $E_2$(TO) (dashed black box), $E_1$(TO) (dashed red box), and $A_1$(LO) (dashed pink box). Figure 2D reveals that, as in the InP pillars, Raman maps corresponding to phonons of different symmetry groups produce spatial profiles with distinct phonon localization.

For a more thorough analysis, we present in Fig. 3A Raman intensity maps of the experimentally observed phonon modes $E_2$(TO), $E_1$(TO), and $A_1$(LO) for two different polarizations ($z(x,x)\bar{z}$ and $z(y,y)\bar{z}$, all four polarization combinations are provided in Supplementary Fig. 9). The leftmost panel of Fig. 3A shows atomic force microscopy (AFM) scans of the SiC grating surface, demonstrating the parabolic appearance of the "crest" which rises above the substrate "trough". As in Fig. 2D, the experimental Raman maps showed distinct differences among the phonon modes: the $E_2$(TO) phonon has pronounced intensity in both crest and trough areas, the $E_1$(TO) phonon mode has intensity concentrated in the crest area, and the $A_1$(LO) phonon mode has intensity concentrated in the trough area. The x-polarized maps generally resemble the y-polarized maps, with only small variations in the field differentiating both configurations (this is largely due to the high-NA objective used in the experiment, as discussed in the supplementary information).

As shown in Fig. 3B we also performed calculations of the time-averaged Raman intensity via Eqs. 1-3, employing the same model used to calculate the Raman response in the InP pillars. Simulations of the SPhP eigenmodes in the grating structure were aided by the AFM measurements of the grating to accurately recreate the surface geometry. The simulations reproduce the spatial variation that is observed experimentally, supporting the observation that the $E_1$ mode is concentrated in the crests and the $A_1$ mode is concentrated in the troughs, while the $E_2$ mode has intensity in both regions. The most notable deviation occurs for the $A_1$ map in the $z(y,y)\bar{z}$ geometry, where the simulation predicts a more uniform polarization than is observed in the experimental map. Further work will be required to reveal why the selection rules applied to the SPhP modes don't reproduce the experimental map in this case. Nevertheless, the generally good agreement between the polarizability coupling theory and the Raman maps indicates that Raman mapping of SPhP modes is achievable when considering a more extensive set of eigenmodes, phonons, and polarization configurations than used to study the InP pillars.

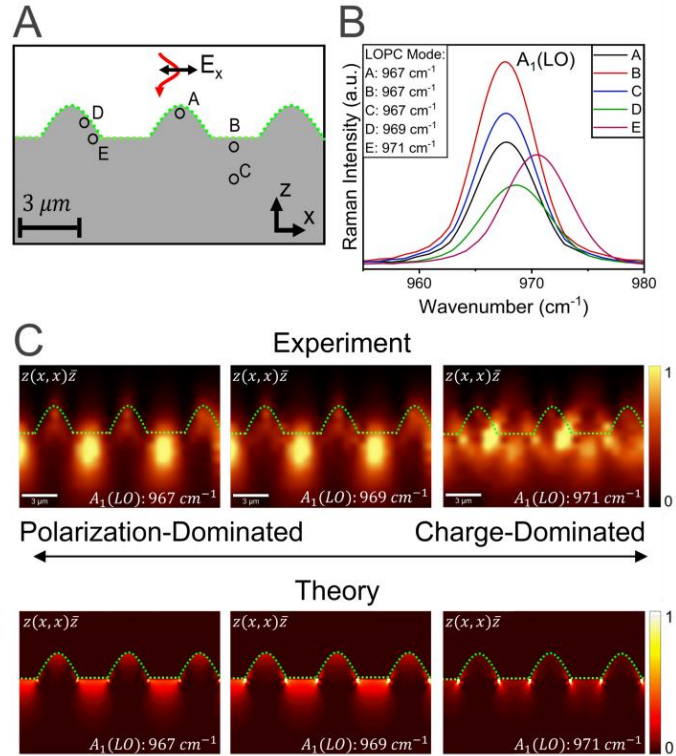

**Figure 4. Raman mapping of electron-phonon coupling in 4H-SiC.** (**A**) Schematic of the Raman mapping investigation of electron phonon-coupling. (**B**) Measured Raman spectra taken from different points on an x-z cross section of the grating, as shown in panel (**A**), in the $z(x,x)\bar{z}$ setup. (**C**) (Top) Experimental and (bottom) theoretical Raman intensity maps of the LO peaks for three spectral positions of the LO peak at 967 cm$^{-1}$, 969 cm$^{-1}$, and 971 cm$^{-1}$.

## Raman Mapping of Electron-Phonon Coupling

While our work supports the remarkable fact that bulk phonons and SPhP modes interact via polarizability coupling, the effects of electron-phonon coupling are also known to have an important impact on the Raman spectra of SPhP modes[4,20,26]. Free carriers interact with phonons via the coupling between the free electron plasma and LO phonons in ionic crystals, known as the longitudinal optical plasmon coupling (LOPC) effect. The LOPC effect is identifiable in Raman spectra via its characteristic blue shifting of the LO mode, and the broadening of the LO peak due to damping from the lossy plasmonic field[39]. The electron-phonon coupling seen through the lineshape of the LO phonon mode is thus a widely used method to investigate the electrical properties of SiC and is frequently employed to estimate the doping density of SiC samples.[39]

This well-established role of Raman spectroscopy in investigating the LOPC effect prompts the idea of using Raman mapping as a way of visualizing the charge distribution induced by the surface modes. In Fig. 4B, we show Raman spectra obtained from different points in the grating structure - as indicated in the cross-sectional view in Fig. 4A. The Raman spectra indicate that the LO mode maintains it spectral position at 967 cm$^{-1}$ for most points in the grating, but near the grating corner (points D and E) the spectral characteristics of the LOPC effect emerge as the LO mode blue shifts to 971 cm$^{-1}$ and broadens. This relatively small shift corresponds to a charge density of $n$ = 5.0×10$^{17}$ cm$^{-3}$, more than two orders of magnitude larger than the estimated doping of the SiC substrate layer (∼ $1.0 \times 10^{15}$ cm$^{-3}$) (See the supplementary materials for more detail on estimating the charge density from the Raman spectra).

We show the experimental Raman maps associated with the LO phonon mode as it shifts from 967 cm$^{-1}$ to 971 cm$^{-1}$ in Fig. 4C. These maps support the observations of Fig. 4B, indicating that as the LO mode blueshifts the Raman signal becomes strongly concentrated in the grating corners due to the corresponding concentration of charge. We compare these maps with theoretical simulations including both the polarizability (following the same selection rules basis as before) and the influence of adding the induced charge (more detail on the LOPC simulations is provided in the supplementary information). The simulations further support experimental observations that increasing densities of free carriers concentrate themselves in the grating corners and thus lead to a spectral shift of the Raman signal in these regions. The observations indicate that confocal Raman microscopy can serve as a powerful tool for visualizing the electron-phonon coupling effects in certain material systems.

## Conclusion

In this work, we demonstrated the three-dimensional mapping of phonon polariton modes in nanopillars of Indium Phosphide and one-dimensional gratings of 4H-Silicon Carbide using visible and near-infrared confocal Raman microscopy. This mapping technique is enabled by the unexpected coupling of bulk phonon modes to the phonon polaritons via changes in the local polarizability, described by the Raman selection rules for each phonon mode. In addition, for $A_1(LO)$ and $E_1(TO)$ in 4H-SiC we observe that free charges also couple to the phonon modes, consistent with the well-known longitudinal optical phonon coupling (LOPC) effect. These observations constitute a method to visualize not only PhP eigenmodes, but also the light-matter interactions between phonons, polaritons, and charge distributions that occur in dielectric nanostructures. This proof-of-principle demonstration is thus promising for the wide range of applications, such as chemical sensing and nonlinear optics, where light-matter interactions are being intensely investigated, as well as in highly anisotropic materials, where volumetric mapping of light-matter interactions is a powerful capability. Furthermore, this work demonstrates a path for extending phonon polaritons physics that have been restricted to wavelengths above 5 μm into the visible regime. For example, the coupling of phonon polaritons to bulk phonons could enable the use of phonon polaritons in applications such surface enhanced Raman scattering (SERS) not previously explored.

## Methods

### Fabrication of InP nanopillars
InP nanopillar samples were prepared in 300×300 μm arrays with varying diameter and pitch. The structures were fabricated on undoped n-type InP, using a SiN hard mask to outline the pillar cross-section. The mask was produced using electron-beam lithography with a fluorine-based inductively coupled plasma etching (ICP). The pillars themselves were then etched into the substrate using a chlorine-based ICP[4]. The InP nanopillars used for this investigation had a diameter of 10 μm, height of 1500 nm, and a center-to-center pitch of 20 μm.

### Fabrication of 4H-SiC gratings
The SiC grating samples were prepared in 200×200 μm arrays with varying pitch and linewidth. The structures were fabricated on an undoped 4H-SiC substrate which had its surface prepared and activated via treatment in the Plasma Preen system. The 1D grating pattern was then generated using maskless direct laser writing using a Nanoscribe 3D printer. The nanostructure pattern is applied to a drop of IP-DIP resin. The 3D printed map is etched into the SiC substrate using a $SF_6/O_2$ reactive ion etching process with $O_2$ being used to remove any remaining polymer resist[41,42]. The SiC gratings used for this investigation have a center-to-center period of 6 μm, a linewidth of 3.32 μm, and a height of 1.41 μm.

### Reflection Spectroscopy
Reflectance measurements for the InP nanopillars were carried out using a Bruker 80v spectrometer, coupled with a Bruker Hyperion 1000 IR microscope. For the SiC gratings, reflectance measurements were carried out using a Bruker Tensor 27 FTIR spectrometer, coupled with a Bruker Hyperion 1000 IR microscope. A reverse-Cassegrain microscope objective was used for both experiments to focus unpolarized IR light onto the sample with an angle of incidence that varied from 12° to 23.6°. The same microscope objective was used to collect the reflected light from the nanostructure sample and subsequently focused onto and detected by a liquid nitrogen-cooled HgCdTe detector. Steady-state reflectance measurements were normalized to the reflectance of a gold mirror. Spectra were taken with an average of 80 scans with 3 cm$^{-1}$ resolution acquired from a 150 μm$^2$ collection area.

**Raman Measurements**

Raman spectra were recorded using a WITec Alpha300R confocal Raman microscope utilizing a charge-coupled device detector. For all experiments, Raman signal was collected using a 100x (0.95 NA) Zeiss EC Epiplan-Neofluar objective. For the investigations on InP, Raman spectra were collected using a frequency-doubled Nd-YAG laser (WITec) providing excitation at an above-bandgap wavelength of 1064 nm perpendicular to the sample surface. For the SiC gratings, a laser excitation at 532 nm was employed. The linear polarization of both excitation lasers was adjusted using a λ/2 calcite waveplate, along with an identical λ/2 waveplate serving as the analyzer. For the InP nanopillars, fifteen 2D (x-y) Raman maps (20 μm by 20 μm, 30 points per line, and 30 lines per image) from different focal planes (1.5 μm) were acquired by automatically scanning the sample along the z-axis (starting with the focal plane above the nanostructure surface and then focusing the beam progressively deeper into the nanostructure in 1.5 μm intervals).

**Electromagnetic Field Simulations**

We used the finite element method (FEM) software COMSOL Multiphysics to simulate the far-field reflection spectrum from both the InP nanopillars and SiC gratings (Fig. 1C and 2A, respectively). For these calculations, the reflection spectrum was calculated for an infinite array of nanostructures with dimensions chosen to match the measured size of both nanostructures. Near-field simulations of the electromagnetic fields (including the polarization fields) are similarly performed using the RF module of COMSOL.

## Acknowledgements (not compulsory)

B.W. acknowledges partial support from the Oklahoma Center for the Advancement of Science and Technology's Research Grant No. AR21-052. C.T.E, C.S.K, M.K., D.C.R, J.G.T. and M.A.M acknowledge funding from the Office of Naval Research. J.G.T. acknowledges support by National Science Foundation under Grant OISE-2230706 and Oklahoma Center for the Advancement of Science and Technology's Research Grant No. AR21-032.

## Author contributions statement

The initial idea was proposed by J.G.T., C.T.E., and D.C.R. The fabricated silicon carbide samples were made by J.G.T., while C.S.K. and M.K. fabricated the indium phosphide samples used in the experiments. The FTIR microscope measurements were carried out by M.A.M., K.E.A., V.R.W., C.T.E. and J.G.T. Microscopy characterization (AFM and SEM) was performed by K.E.A., N.R.S., C.S.K. M.K., B.W. and J.GT. Raman studies were carried out by K.E.A, N.R.S., C.T.E., V.R.W., D.R. and J.G.T. The electromagnetic modeling was performed by K.E.A as well as the analytical theory determining selection rules. All authors contributed and commented on the manuscript with a lead from K.E.A. and J.G.T. The project was supervised by J.G.T.

## Additional information

**Competing interests**

The authors declare no competing financial interests.

# Supplementary Material for

## Mapping Surface Phonon Polaritons with Visible Light

Kiernan E. Arledge, Michael A. Meeker, Chase T. Ellis, Nazli Rasouli Sarabi, Vincent R. Whiteside, Chul Soo Kim, Mijin Kim, Daniel Ratchford, Binbin Weng, and Joseph G. Tischler

*Corresponding author. E-mail: tischler@ou.edu and kiernan.e.arledge-1@ou.edu

**This PDF file includes:**

    Materials and Methods
    Supplementary Text
    Figs. S1 to S11
    References

# Materials and Methods

## Sample preparation

InP nanopillar samples were prepared in 300 x 300 µm arrays with varying diameter and pitch. The structures were fabricated on undoped n-type InP, using a SiN hard mask to outline the pillar cross-section. The mask was produced using electron-beam lithography with a fluorine-based inductively coupled plasma etching (ICP). The pillars themselves were then etched into the substrate using a chlorine-based ICP (1). As mentioned in the main text, the InP nanopillars used for this investigation had a diameter of 10 µm, height of 1500 nm, and a center-to-center pitch of 20 µm. The SiC grating samples were prepared in 200 x 200 µm arrays with varying pitch and linewidth. The structures were fabricated on an undoped 4H-SiC substrate which had its surface prepared and activated via a ozone treatment in the Plasma Preen system. The 1D grating pattern was then generated using maskless direct laser writing using a Nanoscribe 3D printer. The nanostructure pattern is applied to a drop of IP-DIP resin. The 3D printed map is then etched into the SiC substrate using a $SF_6/O_2$ reactive ion etching process with $O_2$ being used to remove any remaining polymer resist (2,3). The SiC gratings used for this investigation have a center-to-center period of 6 µm, a linewidth of 3.32 µm, and a height of 1.41 µm (see supplementary).

## Experiments

**Surface topography measurements:** INP NANOPILLARS SEM IMAGES. Atomic force microscopy (AFM) maps were acquired using an NX10 AFM (Park Systems, South Korea) operating in noncontact mode using a silicon tip (PPP-NCHR; 42 N/m) with a nominal radius of < 10nm. The statistical analysis of the AFM maps was performed using XEI (Park Systems) software. Period, linewidth, and height were measured using a Z-axis cross section for the grating structure.

**Reflection Spectroscopy:** Reflectance measurements for the InP nanopillars were carried out using a Bruker 80v spectrometer, coupled with a Bruker Hyperion 1000 IR microscope. For the

SiC gratings, reflectance measurements were carried out using a Bruker Tensor 27 FT-IR spectrometer, coupled with a Bruker Hyperion IR microscope. A reverse-Cassegrain microscope objective was used for both experiments to focus unpolarized IR light onto the sample with an angle of incidence that varied from 12° to 23.6°. The same microscope objective was used to collect the reflected light from the nanostructure sample and subsequently focused onto and detected by a HgCdTe detector. Steady-state reflectance measurements were normalized to the reflectance of a gold mirror. Spectra were taken with an average of 80 scans with 3 cm$^{-1}$ resolution acquired from a 150 µm$^2$ collection area.

**Raman measurements:** Raman spectra were recorded using a WITec Confocal Raman Microscope system (alpha 300R) with high-speed ultrahigh throughput spectrometers (WITec Inc., Germany). For all experiments, Raman signal was collected using a 100x (0.90 NA) Zeiss EC Epiplan-Neofluar objective (Zeiss, Germany) with a 50 – µm optic fiber, dispersed by a diffraction grating of 600 grooves/mm, and collected on a EMCCD detector. For the investigations on InP, Raman spectra were collected using a frequency-doubled Nd-YAG laser (WITec) providing excitation at an above-bandgap wavelength of 1064 nm perpendicular to the sample surface. For the SiC gratings, a laser excitation at 532 nm was employed. The linear polarization of both excitation lasers was adjusted using a λ/2 calcite waveplate, along with an identical λ/2 waveplate serving as the analyzer. Rayleigh scattered light was blocked by an edge filter. For the InP experiments, laser power was set to the 15 mW range and satisfactory signal-to-noise ratio was achieved using an acquisition time of 12 s. For the SiC experiments, laser power was set to the 30 mW range and satisfactory signal-to-noise ratio was achieved using an acquisition time of 7.5 s. For the InP nanopillars, fifteen confocal 2D (*x-y*) Raman maps (20 µm by 20 µm, 30 points per line, and 30 lines per image) from different focal planes (1.5 µm) were acquired at high spectral and spatial resolution by automatically scanning the sample along the z-axis (starting with the focal plane above the nanostructure surface and then focusing the beam progressively deeper into the nanostructure in 1.5 μm intervals). Data acquisition, evaluation, and processing were performed using the WITec Project Management and Image Project Plus software.

## Data Analysis

**Raman Data Processing:** All Raman spectra shown in the main article were fitted to single-peak Gaussian functions. The center wavelength of each Gaussian is used to obtain the phonon energies reported.

## Simulation

**Electromagnetic Field Simulations:** We used the finite element method (FEM) software COMSOL Multiphysics to simulate the far-field reflection spectrum from both the InP nanopillars and SiC gratings (Fig. 1C and 2A, respectively). For these calculations, the reflection spectrum was calculated for an infinite array of the nanostructures with dimensions chosen to match the measured size of both nanostructures. Near-field simulations of the electromagnetic fields (including the polarization fields) are similarly performed using the RF module of COMSOL. Simulations are performed for incident angles of 0°, 32°, and 64°, and the optical fields shown (Fig. 1E,F; Fig. 3, and Fig 4C) consist of equally weighted maps for each angle.

**Electron-Phonon Coupling Simulations:** The process of determining the spatial profile of electron-phonon coupling, shown in Fig. 4C, requires simulating the distribution of induced charges in the SiC gratings. Volume charge distributions (at point **r** and frequency ω) in the nanostructure are obtained, following the approach of Marty (4,5), as the divergence of the polarization vector:

$$\rho(\boldsymbol{r},\omega) = -\nabla \cdot \boldsymbol{P}(\boldsymbol{r},\omega)$$

Maps of the charge distribution measured are reconstructed in the same manner as the polarization fields, as described in the next section. In order to estimate the amount of induced charge in the nanostructures, the shift in the Raman spectra is used to approximately determine the induced charge density. As mentioned in the main text, in the Raman spectrum of polar dielectrics like SiC, the spectral position of the LO phonon peak will blueshift and broaden its full-width half maximum when the dielectric is doped with free carriers (the LOPC effect). This effect occurs because the large dipole moment of the LO phonon can couple to the electron plasma (an effect which becomes more prominent as the doping density increases). Using the Raman spectra, the carrier density of 4H-SiC in the low doping regime is estimated using the approach of Nakashima, where the blueshift of the LO mode, $\Delta\omega$, is related to the carrier density $n$ via a simple expression (6):

$$n = 1.25 \times 10^{17} \Delta\omega.$$

For the bulk SiC substrate there is no measured blueshift, indicated the very low background doping. Nevertheless, in very concentrated positions in the grating structure, a blueshift of up to $\Delta\omega = 5\ cm^{-1}$ is observed in the Raman spectra (Fig. 4B). To test the hypothesis that the local blueshifting is due to the electron-phonon coupling effect, we simulated both the polarization and charge distribution in the 1D gratings. Using Eq. (#) above, we can estimate the induced charge in the nanostructures from the spectral shift in the LO phonon at that location. To obtain the maps shown in Fig. 4C, the polarization field of the SPhP modes based on the Raman tensor are computed (identical to Fig. 3 for the LO mode), which serve as the baseline for spectra with no blueshifting. For larger carrier densities, the induced charge is computed using Eq. S1, and the final maps are a combination of both polarization and charge maps.

## Reconstruction

The process to reconstruct the maps of SPhP modes (Fig. 1D,E, Fig. 3, Fig. 4C) generally follows the logical approach of EELS reconstruction (7-9). The details of the Raman reconstruction method are discussed in the supplementary materials. Procedurally, the Raman maps are constructed using an eigenmode decomposition, as in EELS, where the overall Raman scattering probability for a given phonon mode can be written as:

$$\mathcal{P}_{Raman} = \sum_k C_k \mathcal{P}_{Raman}(u_k)$$

where the sum is performed over the index $k$, associated with a geometric SPhP eigenmode $u_k$ (such as a monopole in the nanopillars), $C_k$ is a coefficient specific to each geometric eigenmode and $\mathcal{P}_{Raman}(u_k)$ is the Raman scattering probability generated from each SPhP eigenmode, calculated from the Raman selection rules (see supplementary). Unlike most EELS experiments, we do not seek to algorithmically obtain $u_k$ and $C_k$ that provide the best match to experimental data, although that will likely be a future direction when theoretical knowledge of Raman mapping of SPhP is further investigated. Rather, in our work we obtain the eigenmodes $u_k$ from the reflection spectrum, sharp dips are identified as eigenmodes and constitute the basis set. Additionally, for simplicity, we arbitrarily set all coefficients $C_k$ to be equivalent (1, in this case), so that each simulated Raman map is an admixture of the Raman scattering probability associated with each SPhP mode. We anticipate that if even more precise methods are used to obtain the eigenbasis and coefficients, then the matching between experiment and simulations might be improved substantially, particularly the effect of the substrate.

# Supplementary text

## Supplementary experimental data

### InP Nanopillars

The geometry of the InP pillar nanostructures was estimated from SEM imaging. The SEM images in Fig. S1 a,b demonstrate two periodic arrays of nanopillars with different characteristic dimensions. Fig. S1a shows the SEM image of the nanopillar array used in this paper's investigation of Raman maps, with a measurement of the top diameter overlaid. Fig. S1b shows an array of nanopillars with smaller diameter, which will be investigated in the following supplementary data.

Both nanopillar arrays support localized SPhP modes, as demonstrated in Fig. S1c,d, which give simulated reflection spectra using the full-wave, 3D electromagnetic simulation approach in COMSOL described in the Materials and Methods. For both nanopillar sizes comparing the off-array reflection spectrum (black curve) with the on-array reflection spectrum (red curve) reveals that the nanopillars induce reflection dips associated with localized SPhP modes.

The SPhP modes of both nanopillar sizes can be mapped using Raman imaging techniques, as shown in Fig. S2. Figure S2 shows x-y Raman image cuts for four different nanopillar arrays, showing both the TO and LO mode for each nanopillar size selected. The image size is fixed at 20 μm x 20 μm.

The uppermost array in Fig. S2 corresponds to an array with a period of 20 μm and a diameter of 10 μm, which corresponds to the Raman images in the paper. As discussed in the paper, the correspondence between the TO(LO) phonon modes and the dipole(monopole) SPhP modes is evident. The panel below (period of 16 μm and diameter of 8 μm) shows similar images of the localized modes. The next panel below (period of 10 μm and diameter of 5 μm) shows multiple nanopillars within the array. While the obtained maps have become blurrier, the dipole and monopole field distributions remain evident and appear to be similar in each pillar of the array. The bottom panel (period of 6 μm and diameter of 3 μm) shows a map containing almost 12 full pillars. The persistence of monopole and dipole fields appears to be discernible, but the small size has made imaging difficult given the diffraction limitation. Nevertheless, the presence of localized modes can be verified for nanopillar sizes far smaller than the SPhP wavelength (~30 μm) and is not isolated to larger nanostructures. The largest nanopillars are chosen for the paper since their images are the easiest to visualize.

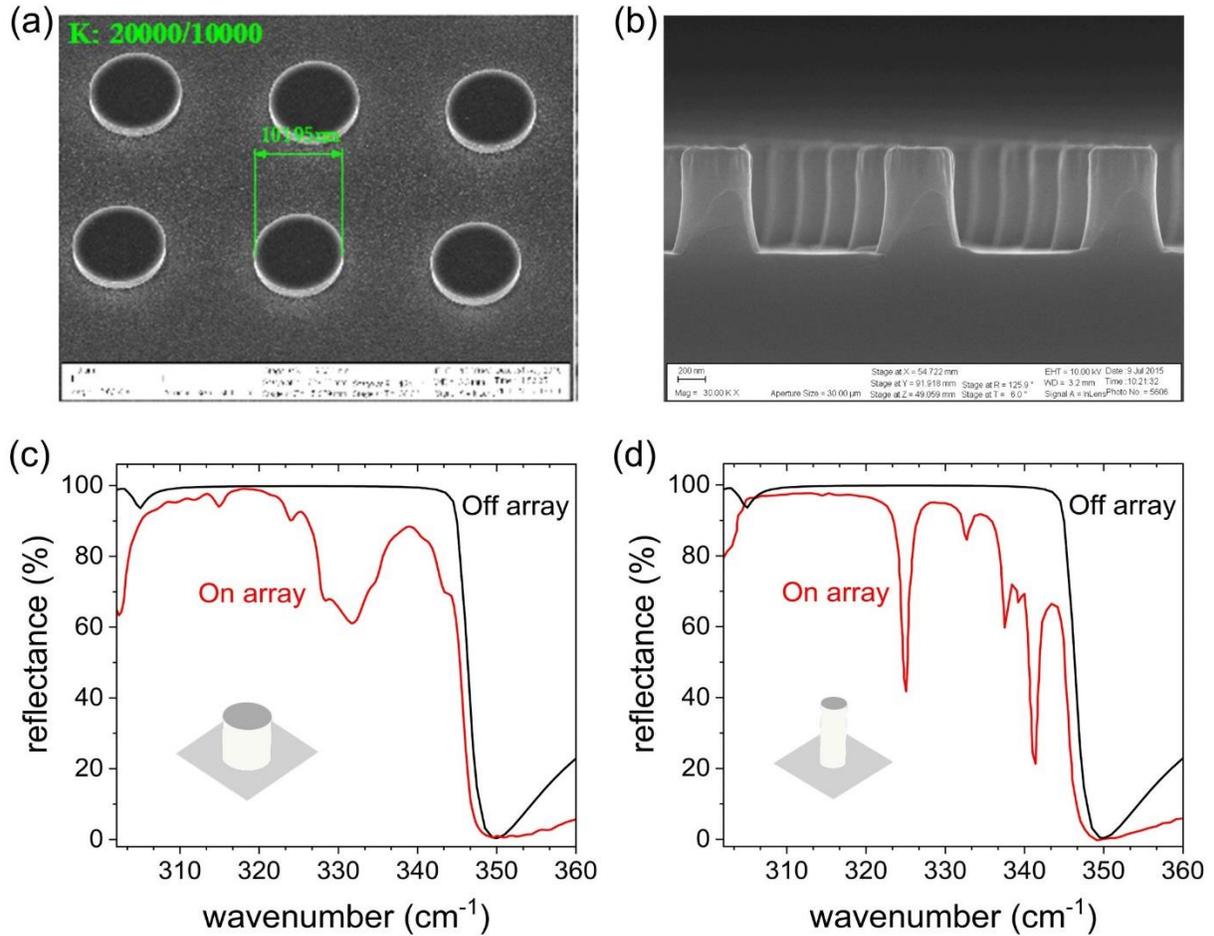

Fig. S1: SPhP resonances in InP nanopillars. A. SEM image of the InP nanopillars with 10 µm diameter, used in the paper. B. Cross-sectional SEM image of a nanopillar array with 300 nm diameters, far below the free-space polariton wavelength. C. Off array (black) and On-array (red) reflection spectra for the larger pillar array shown in panel (a) demonstrating the resonances associated with the localized SPhP modes. D. Off array (black) and on array (red) reflection spectra for the smaller pillar array shown in panel (b) demonstrating localized SPhP resonances.

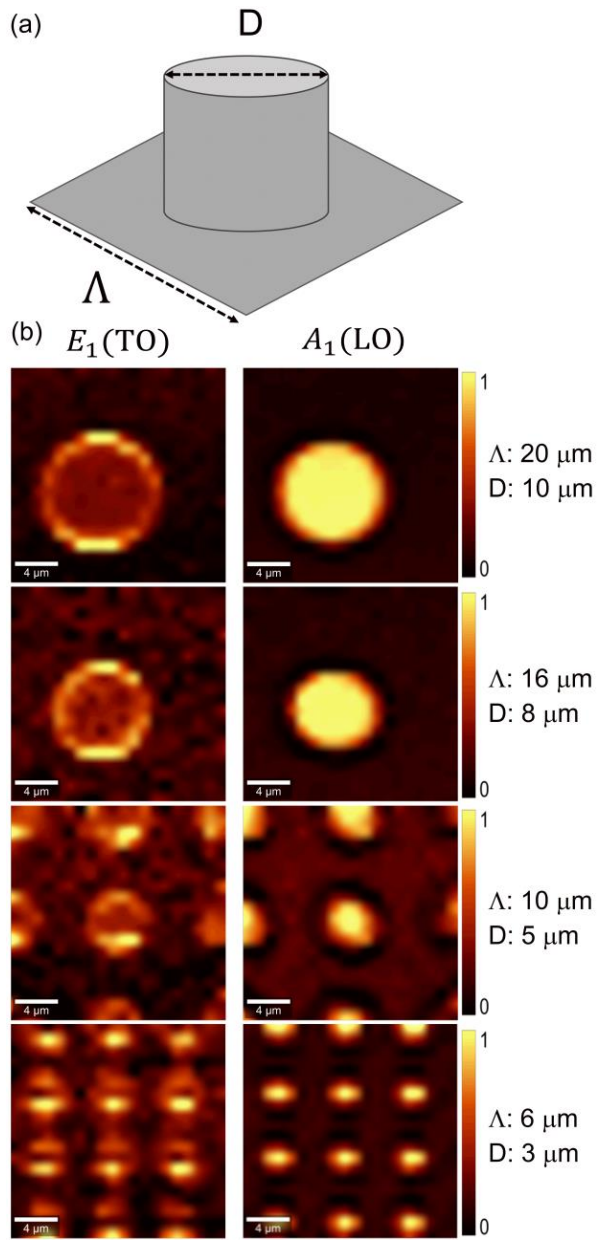

Fig. S2: Raman mapping for different nanopillar arrays. A. Sketch of the nanopillar geometry, with diameter (D) and period (Λ) shown. B. Raman images of the TO and LO phonon modes for nanopillars of different sizes. The top panel shows a Λ = 20 μm, D = 10 μm nanopillar, as shown in the paper. Lower panels show nanopillars of decreasing feature sizes, demonstrating the continued confinement of the optical mode, with limited imaging resolution.

Figure 1 in the paper gives both x-y and x-z Raman maps for the largest nanopillar array, but x-y Raman scans were also collected for various focal depths in the order to visualize the mode intensity in the nanopillar and within the InP substrate, as depicted schematically in Figure S3 (a). As Figure S3(a) shows, we start at "Stack A" collected roughly 3.5 µm above the InP substrate and scan progressively down to "Stack J" roughly 1.5 µm into the substrate. Figure S3(b) demonstrates the result of these scans for the TO and LO phonons in the 10 µm diameter nanopillar. As the scans focus deeper into the InP substrate, we can visualize the monopole-like and dipole-like modes evolve through the vertical direction. In particular, the dipole-like TO phonon map has its profile fall apart and become noisier, while the monopole-like mode retains a strong profile through Stack J.

The results of the stack scan reinforce Figure 1(D) from the paper, where the cross-sectional map shows the TO mode losing its dipole-like field profile ~1.5 µm deep into the substrate, while the monopole remains strong over a larger distance. The results of these stack scans are also largely consistent with other investigations of SPhPs in nanopillars, which have noted the strong interaction of the monopole mode with the material substrate, necessary to maintain its charge neutrality.

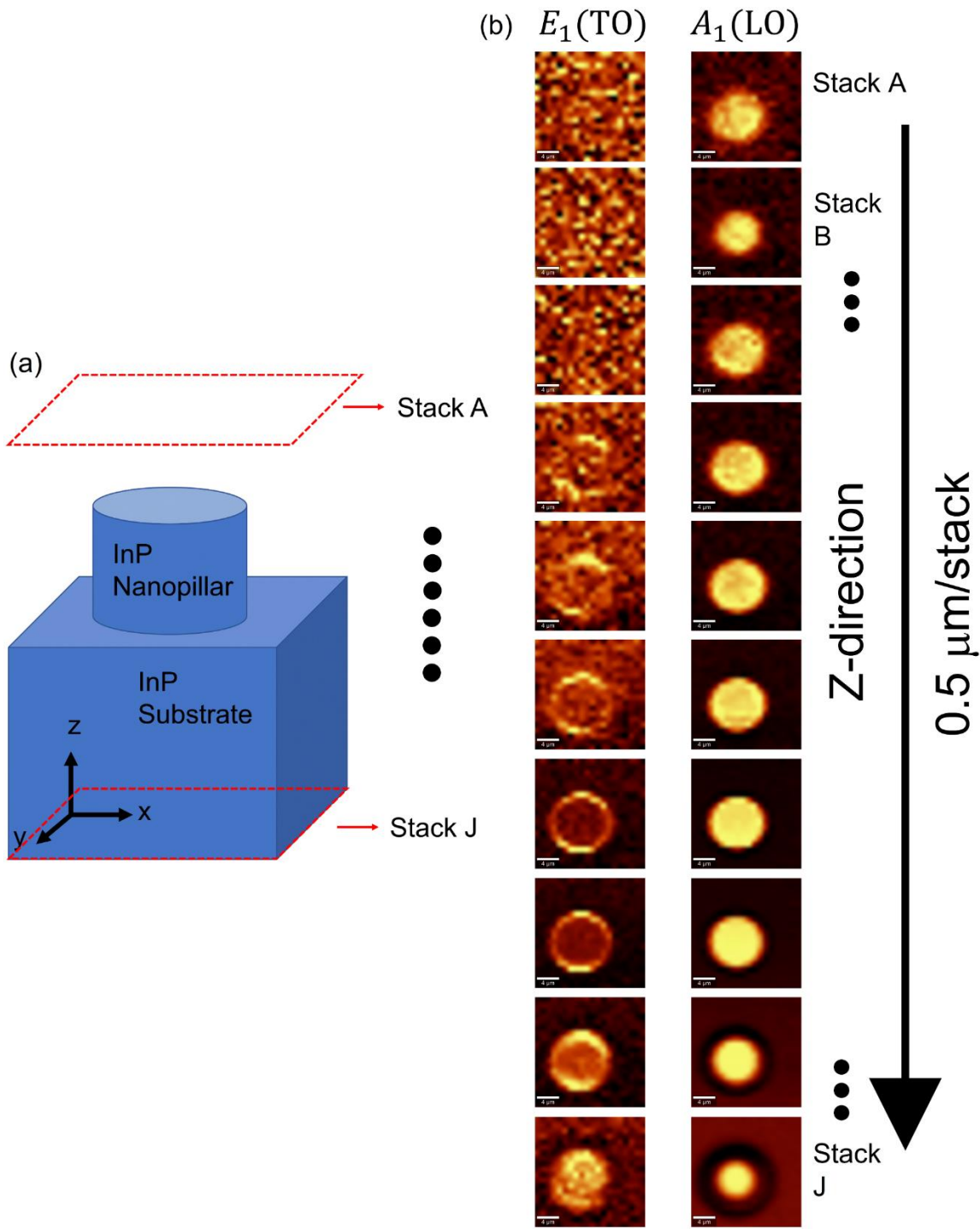

Fig. S3: Stack Scan of InP nanopillars. (A). Sketch of the experiment, where Raman maps in the x-y plane are collected at a constant z-focus, generating a stack of Raman maps from A to J, focusing progressively deeper. (B) Results of the stack scan measurements for the $E_1$(TO) and $A_1$(LO) phonon modes. The monopole-like LO mode has a similar field profile for almost the entire stack

scan but the dipole-like TO mode profile becomes very noisy and more monopole-like as it progresses deeper into the substrate.

## One dimensional (1D) SiC Gratings

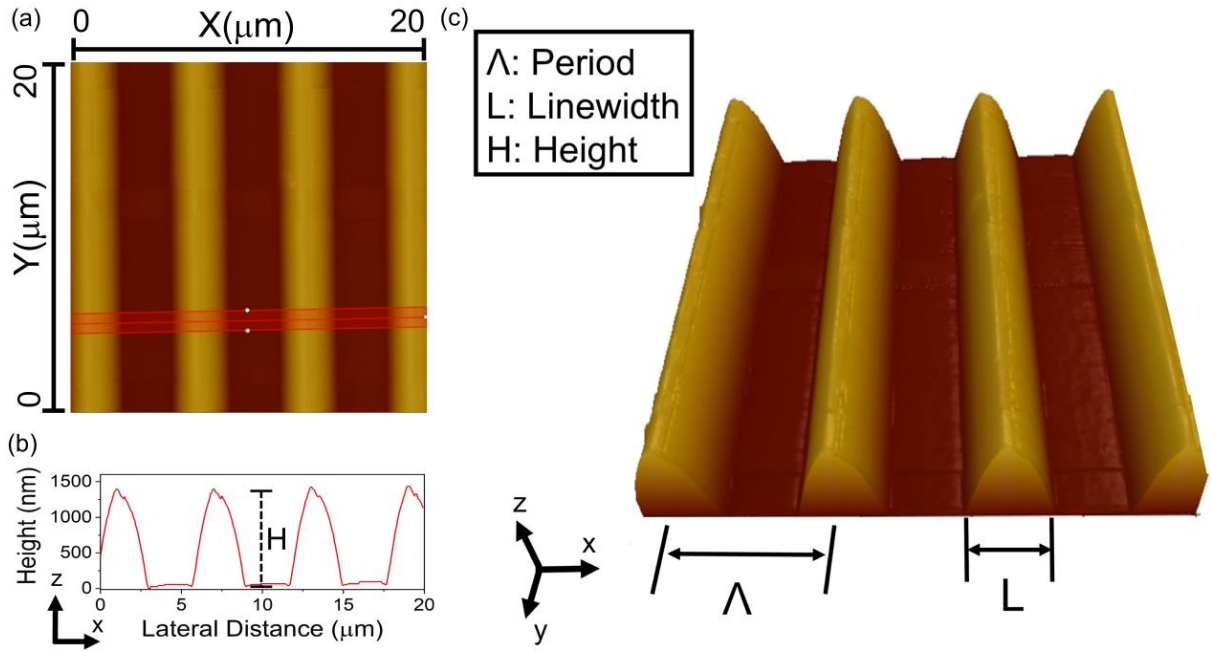

Fig. S4: AFM measurements of SiC gratings. (A). x-y AFM scan for 1D grating used in the article, the red line shows the cut where z values are averaged to produce a cross-sectional profile. (B) cross-sectional profile of grating structure in the x-z plane, used to extract all geometric parameters used in modelling. The height H is shown. (C). 3D view of the gratings which gives a clear view of the parabolic geometry produced by the etching process. The period Λ and linewidth L are shown.

The geometry of the SiC gratings were measured using non-contact atomic force microscopy (AFM). Figure S4 shows the results of the AFM measurements on the grating array used in the article. Fig. S4(a) shows a x-y view of the AFM showing the location of the y-cut which is used to measure the grating structural parameters. The resulting x-z linescan, shown in Fig. S4(b), provides a cross-sectional view of the grating. The AFM scans show the parabolic shape of the gratings, which is a result of the etching process employed on SiC. The cross-sectional scan also allows for a measurement of the average grating height H above the substrate, along with the average period Λ and bottom linewidth L, as shown in the full 3D scan in Fig. S4(c). With Λ,L, and H having been measured, we can use those parameters to obtain a parabolic fit for the grating structure of the form z = ax$^2$+bx+c, determining the value for parameters a,b, and c. These fits are then used to recreate the grating geometry used to simulate the SPhP modes in Fig. 2(a) and Fig. S5.

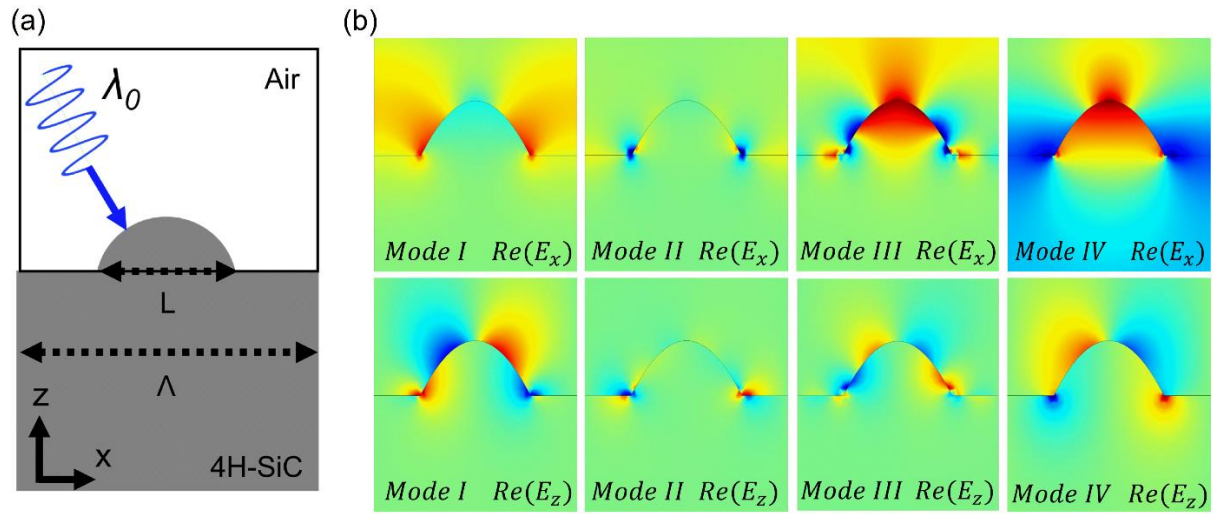

Fig. S5: FEM simulations of SPhP modes in SiC gratings. (A). Diagram of the computational geometry used for the modelling. The period Λ and linewidth L obtained in Fig. S4 are used to create the structure. (B). The Re($E_x$) (top) and Re($E_z$) (bottom) distributions of the four SPhP resonances in the SiC gratings. Energies of modes I-IV are shown in Fig. 2(a) in the article.

Having determined the grating geometry from AFM measurements, we can create accurate FEM models for the nanostructures in the RF package of COMSOL Multiphysics. Figure 2 in the article shows the results of FTIR measurements of the grating structure compared to the computational model which is designed based on the geometric measurements in Fig. S4. As mentioned in the article, the localized SPhP modes in Fig. 2 do not have the symmetry seen in the InP nanopillars, so we can't simply classify them as monopole modes or dipole modes. Lacking a straightforward naming scheme, we labeled the modes I,II,III, and IV in Figure 2(a).

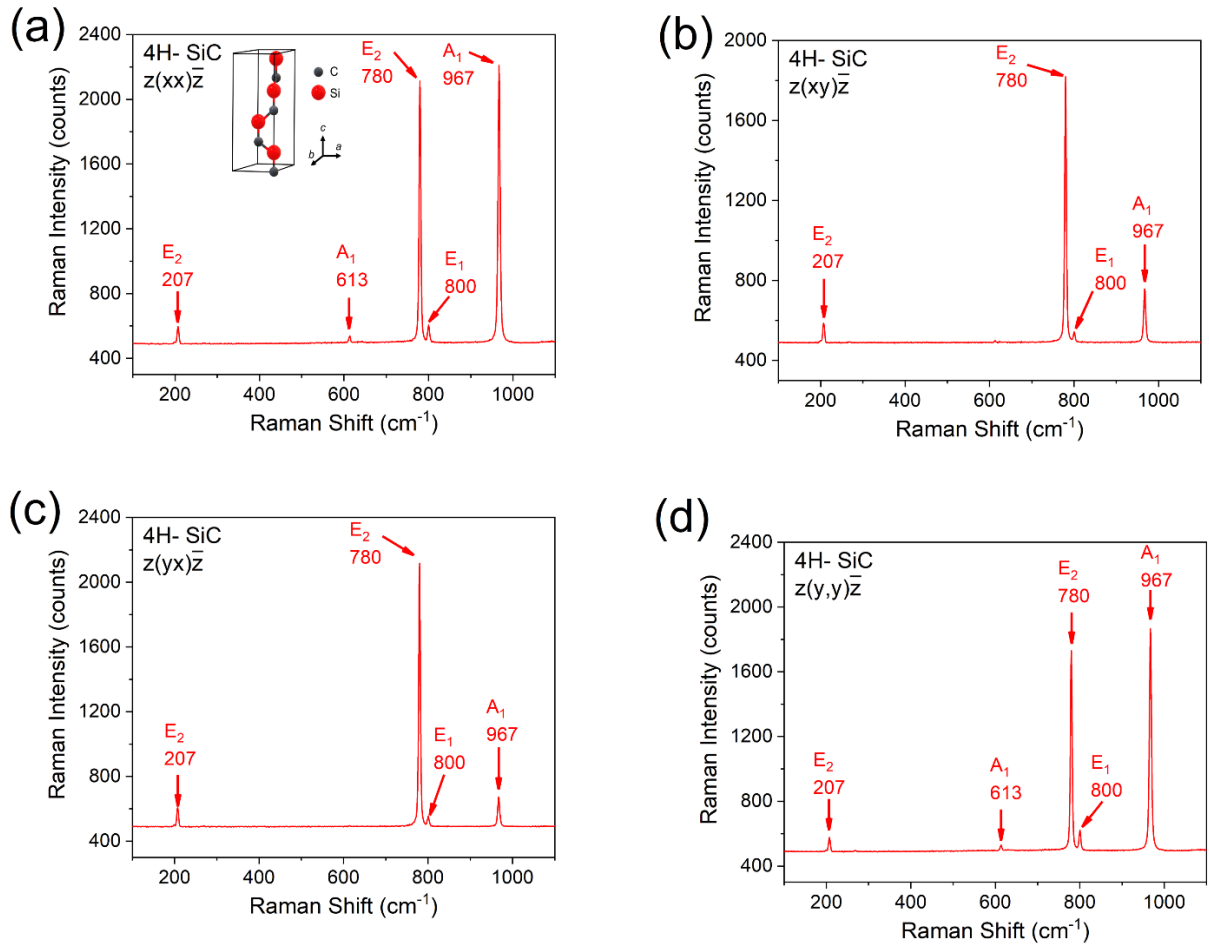

Fig. S6: Backscattering Raman spectra of 4H-SiC. We present the backscattering Raman spectra for bulk SiC material in all four polarizer-analyzer configurations: $z(xx)\bar{z}$ (a), $z(xy)\bar{z}$ (b), $z(yx)\bar{z}$ (c), and $z(yy)\bar{z}$ (d). Phonon modes are labeled by their symmetry classification.

In Figure S5 we show the electromagnetic fields of the SPhP modes shown in Figure 2(a).

Figure 5S(a) presents a schematic of the computational unit cell. The 4H-SiC grating structure is built using period Λ, linewidth L, and height H as determined in Fig. 4S. The structure is illuminated by light of wavelength $\lambda_0$, which is then swept to obtain the reflection spectrum. For each of the modes I, II, III, and IV labeled in Figure 2(a), we show the $E_x$ and $E_z$ fields in the nanostructure, demonstrating the subdiffractional confinement of electromagnetic energy by the SPhP modes. These 4 modes were selected to serve as the basis for the eigenmode reconstruction used to produce the simulated Raman maps in Fig. 3(b) and Fig. 4(b).

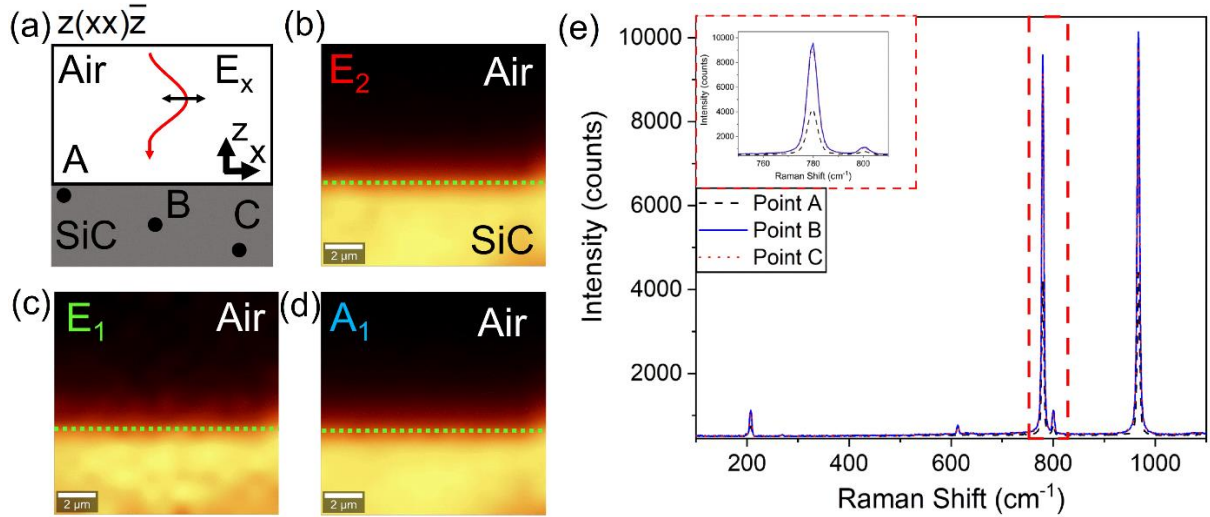

Fig. S7: Raman maps of bulk 4H-SiC. (A) Diagram of the experimental measurement. The incoming light is x-polarized incident on the SiC-air interface. Experimental maps of the Raman intensity are obtained for the $E_2$ (panel B), $E_1$ (panel C), and $A_1$ (panel D) phonon modes. (E). Collected Raman spectra at the three points shown in panel (A), the intensities get stronger with increasing focus depth, but relative intensities remain constant.

In Fig. 2(b) in the article we show the $z(x,x)\bar{z}$ Raman spectrum for the 4H-SiC substrate in order to present the 6 Raman-active phonon modes measured. Since all four polarizer and analyzer combinations were used to obtain Raman maps of the SPhP modes, we measured the bulk Raman spectrum of 4H-SiC under all four polarizer-analyzer combinations as a reference. Figure S5 shows the measured Raman spectra for the $z(x,x)\bar{z}$ (a), $z(x,y)\bar{z}$ (b), $z(y,x)\bar{z}$ (c), and $z(y,y)\bar{z}$ (d) scattering geometries. These Raman scans show that both the parallel (xx and yy) and cross (xy and yx) have similarities. In particular, the parallel polarizations show all 6 Raman-active phonons with similar relative intensities. The cross polarizations lead to considerable decreases in the $E_1$ and $A_1$ phonon intensities, as predicted by the Raman selection rules.

In addition to measuring the bulk Raman scattering in all polarizations, we also performed Raman mapping in the bulk SiC substrate. Figure S7(a) and Figure S8(a) show schematic representations of the SiC scattering geometries for both $z(x,x)\bar{z}$ (S7) and $z(x,x)\bar{z}$ (S8) polarization configurations. Using these setups, panels (b),(c), and (d) show the obtained Raman maps for all three optical phonon modes, $E_2$ (red), $E_1$ (green), and $A_1$ (blue). While there are variations in the signal-to-noise ratio among the phonon modes, all the Raman maps display the expected isotropy with uniform distribution, which distinguishes the bulk material from nanostructured 4H-SiC. Finally, Figure S7 (e) and Figure S8 (e) show the Raman spectra obtained at points A,B, and C in Figure 7,8 (a). There is expected depth dependence to the intensity of the Raman signals, but the relative intensities remain constant for all phonon modes, in contrast to the Raman signals in the SiC gratings where the relative phonon intensities show high local variation (Fig. 4A, for example).

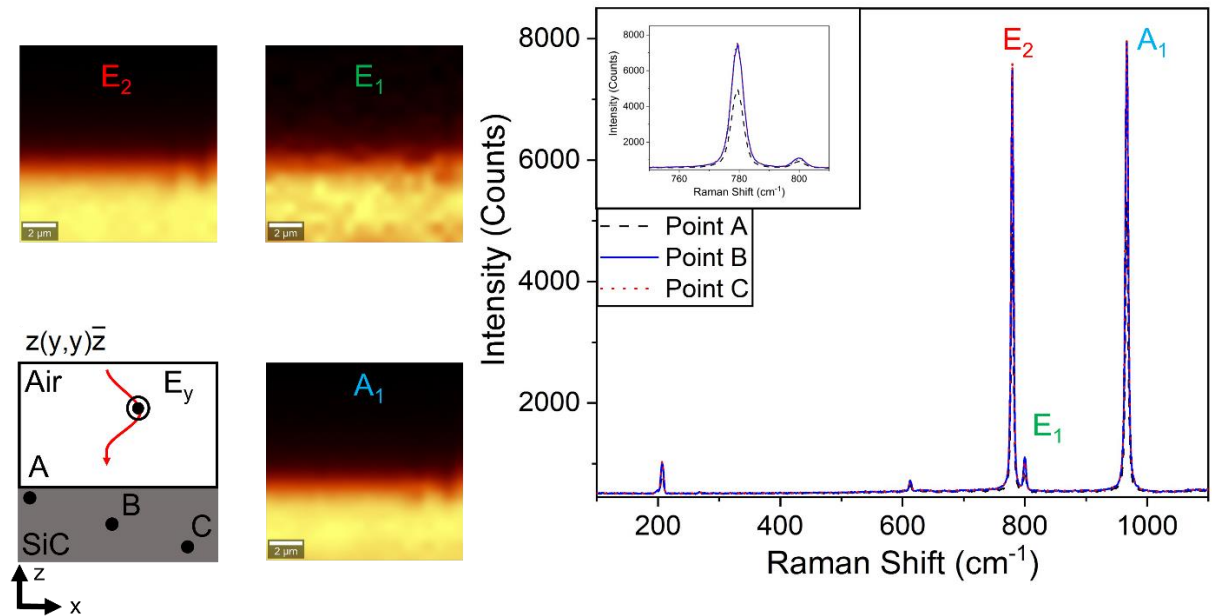

Fig. S8: Raman maps of bulk 4H-SiC. (A) Diagram of the experimental measurement. The incoming light is y-polarized incident on the SiC-air interface. Experimental maps of the Raman intensity are obtained for the $E_2$ (panel B), $E_1$ (panel C), and $A_1$ (panel D) phonon modes. (E). Collected Raman spectra at the three points shown in panel (A), the intensities get stronger with increasing focus depth, but relative intensities remain constant.

As a complement to Fig. S5, in Fig. S9 we show the results of the Raman maps obtained for the three optical phonon modes in all four polarizer configurations. The left most boxes denote the phonon symmetry which all modes in that row correspond to: $E_1$ (green), $E_2$ (red), and $A_1$ (blue). The top panels denote the polarizer-analyzer configuration corresponding to its column and additionally outlines the grating geometry with dashed green lines. As in Fig. 3, the Raman modes are largely similar across all the polarization configurations. Because the phonon modes mix polarizabilities in different directions, as the angle of incidence increases the relative strengths of the polarizabilities changes with the incident electric field. Thus, the large NA of our microscope objective acts to "smear" the Raman polarizabilities in both x and y polarizations, for a lower NA with a smaller solid angle the selection rules might be more visually evident.

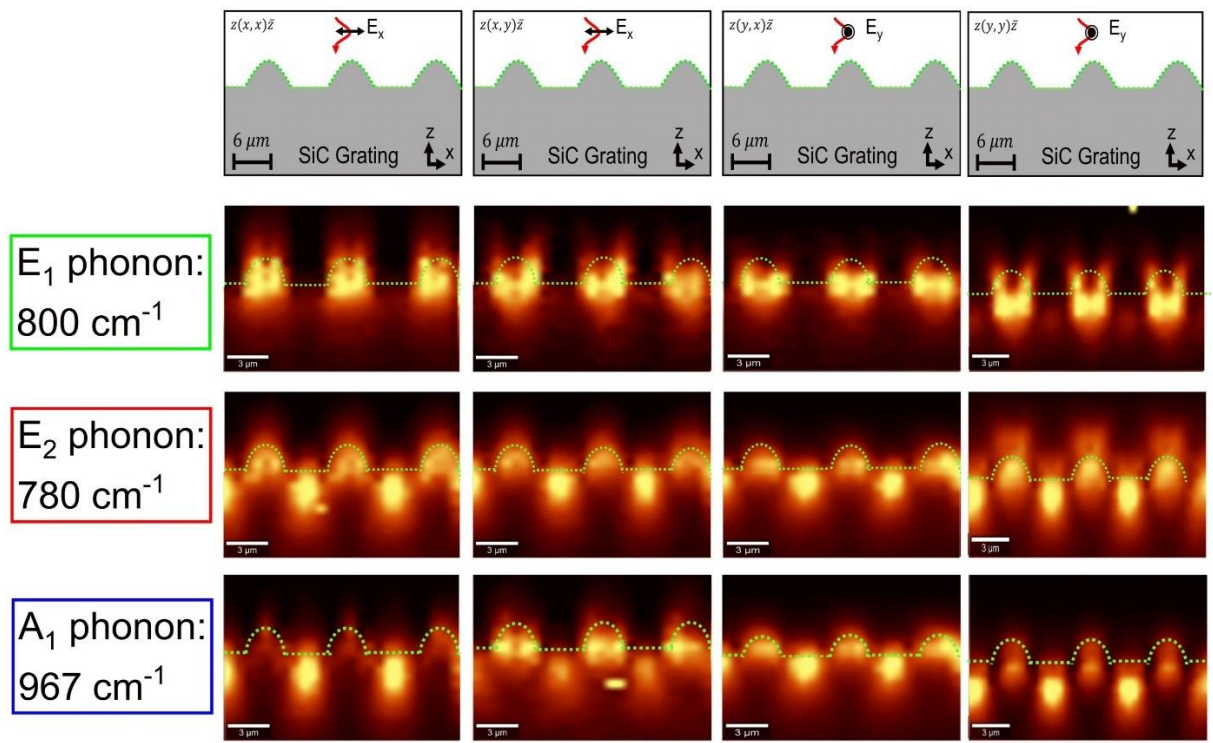

Fig. S9: Collected Raman maps in all polarization configurations for the optical phonon modes in the 4H-SiC grating nanostructures. The top row denotes the geometric setup for the Raman experiment, showing all four polarizer and analyzer configurations. The left column gives the three optical phonon modes. These 12 maps comprise all the polarized Raman maps obtained in the backscattering configuration.

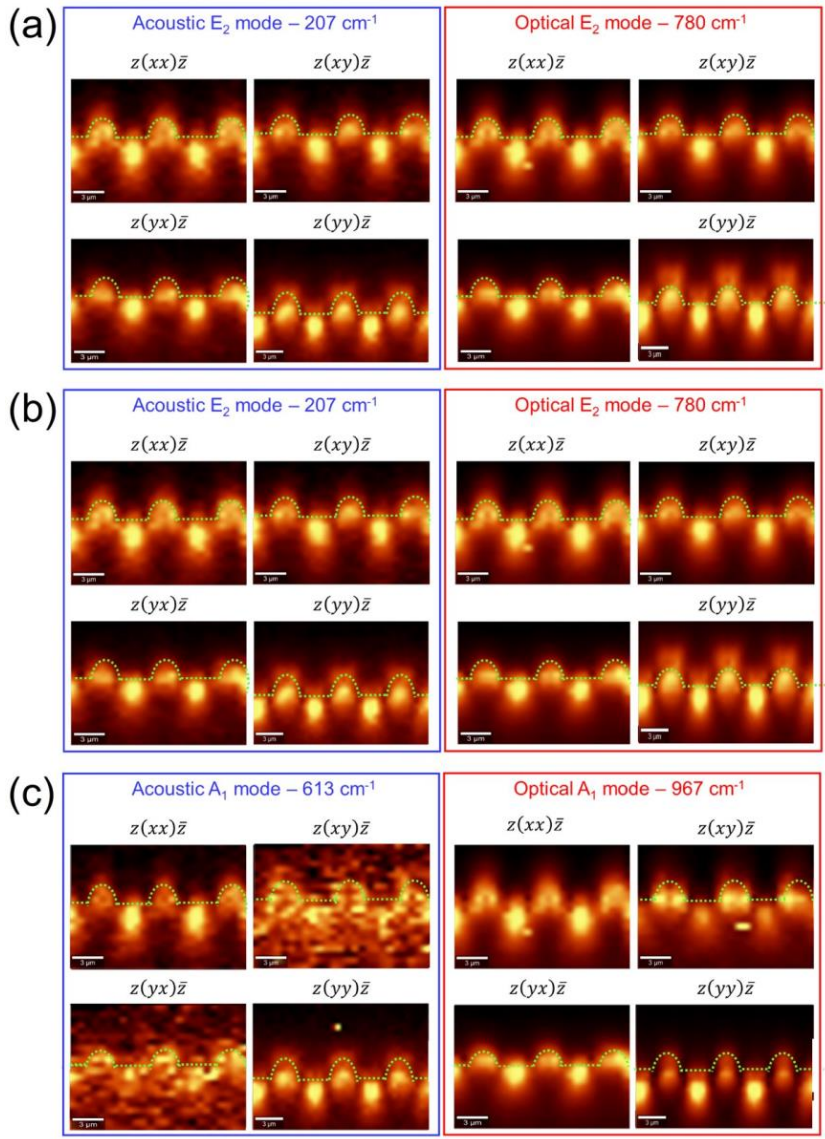

Fig. S10: Acoustic Phonon Raman Maps. (A). Collected Raman maps in all polarizer-analyzer configurations for both the acoustic and optical phonon mode with $E_2$ group symmetry, both sets of maps show remarkable agreement. (B). Same as (A) for the $E_2$ symmetry group phonons. (C). Same as (A) for the $A_1$ symmetry group phonons.

While the paper focuses entirely on the optical phonon modes in 4H-SiC and on their coupling to the SPhP modes, we were also able to obtain Raman maps for the three acoustic phonon modes, as shown in Fig. 2A and Fig. S6. Figure S10 displays the obtained Raman maps in all polarizer configurations, specifically to compare the optical and acoustic phonon modes for each symmetry classification. As shown in Fig. 2(b) and Fig. S5, there is one optical phonon and one acoustic phonon each for all three backscattering Raman-active phonon symmetries in 4H-SiC. Figure S10 shows the Raman maps for both the acoustic and optical phonon modes in the SiC grating structures for the $E_2$ group symmetry (panel a), $E_1$ group symmetry (panel b), and $A_1$ group symmetry (panel c). The Raman maps of both the acoustic and optical phonon clearly depict similar field distributions in the SiC gratings. Thus, the Raman maps shown in Fig. S10 demonstrate the remarkable fact that the coupling between SPhPs and bulk phonon modes applies to the acoustic phonon modes, as well. That is, even though acoustic phonons are thought to play no role in the physics of SPhPs and little role in optical systems in general, our Raman study provides evidence that their interaction with the SPhP modes mirrors that of the optical phonon modes owing to the small Raman polarizability associated with acoustic phonons. This surprising discovery indicates that there is a wider range of material interactions related to SPhp modes than previously thought, including material interactions far outside the Reststrahlen band.